\documentclass[twocolumn,superscriptaddress,floatfix,preprintnumbers,amssymb ,amsmath]{revtex4}
\usepackage{graphicx}
\usepackage{dcolumn}
\usepackage{bm}
\usepackage[latin1]{inputenc}
\usepackage[mathscr]{eucal}
\usepackage{epsfig}
\usepackage{rotating}

\begin{document}

\title{Normal metal - superconductor tunnel junction as a Brownian refrigerator}

\author{J.P. Pekola}
\affiliation{Low Temperature Laboratory, Helsinki University of
Technology, P.O. Box 3500, 02015 TKK, Finland}
\author{F.W.J. Hekking}
\affiliation{ Laboratoire de Physique et Mod\'elisation des Milieux
Condens\'es, C.N.R.S. and Universit\'e Joseph Fourier, B.P. 166,
38042 Grenoble Cedex 9, France}


\begin{abstract}
Thermal noise generated by a hot resistor (resistance $R$) can,
under proper conditions, catalyze heat removal from a cold normal
metal (N) in contact with a superconductor (S) via a tunnel barrier.
Such a NIS junction acts as Maxwell's demon, rectifying the heat
flow. Upon reversal of the temperature gradient between the resistor
and the junction the heat fluxes are reversed: this presents a
regime which is not accessible in an ordinary voltage-biased NIS
structure. We obtain analytical results for the cooling performance
in an idealized high impedance environment, and perform numerical
calculations for general $R$. We conclude by assessing the
experimental feasibility of the proposed effect.
\end{abstract}

\maketitle

In 1867, Maxwell suggested a demon that attempts to violate the second law of
thermodynamics \cite{Leff}. The demon acts between two containers A and B, initially at
the same temperature; it exclusively allows hot particles to pass from container A to
container B, and cold ones from B to A. This process would lead to a decrease of entropy
if the system were isolated, and could then be used for useful work. Ever since, this
thought experiment has intrigued physicists, see, e.g., Refs.
\cite{Leff,landauer61,bennett82,zurek03}. Yet the demon needs to exchange energy with the
containers in order to function properly. Thereby the entropy of the {\em whole} system
(including the demon) is always increasing, rendering thermodynamics intact.

\begin{figure}
\begin{center}
\includegraphics[width=0.5
\textwidth]{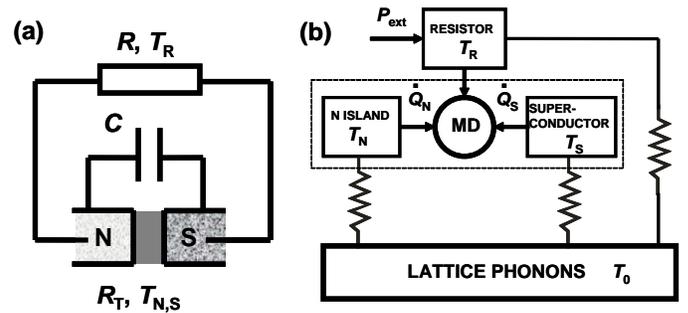} \caption{Schematic presentation of the system.
In (a) we show the electrical diagram of the resistor at temperature
$T_{\rm R}$, and the tunnel junction. The parallel capacitance $C$
includes that of the junction and a possible shunt capacitor. In (b)
we show the thermal diagram of the system. The NIS tunnel junction
acts as a Brownian heat engine, or as Maxwell's demon (MD) between N
and S. $\dot{Q}_{\rm N}$ and $\dot{Q}_{\rm S}$ are the heat fluxes
out from N and S, respectively, and $P_{\rm ext}$ denotes the
external power needed to create the temperature bias of the
resistor. The thermal resistances indicate phonon coupling of the
electrical subsystems.}\label{fig:Fig1}
\end{center}
\end{figure}

In this Letter we present a particularly illustrative example of Maxwell's demon, which
can be realized experimentally in a straightforward way. Our system is a Brownian
refrigerator \cite{broeck06} in close analogy to Brownian motors and thermal ratchets
\cite{buttiker87,parrondo02,reimann02,astumian02,sokolov98}. It conveys heat
unidirectionally in response to random noise. Specifically, we consider a tunnel junction
between a normal metal and a superconductor (NIS junction) subject to the thermal noise
of a resistor at temperature $T_{\rm R}$, see Fig.~\ref{fig:Fig1}. The temperatures of
the electrodes N and S are $T_{\rm N}$ and $T_{\rm S}$, respectively. The capacitance $C$
consists of that of the junction itself and the surrounding circuit. The resistor and the
junction can be connected by superconducting lines which efficiently suppress electronic
thermal conductance, and thus enable us to discuss solely the photonic heat exchange via
the lines as in Refs. \cite{schmidt04,meschke06}. The N side can be connected to the
superconducting line via a metal-to-metal SN contact, which provides perfect electrical
transmission but, due to Andreev reflection, prevents heat flow
\cite{andreev64,giazotto06}. We note already here that the presented NIS structure can be
replaced by a more conventional symmetric SINIS device with two tunnel junctions
back-to-back. This will be discussed towards the end of the present Letter. We also note
that the idealized circuit of Fig. \ref{fig:Fig1} (a) as well as the thermal model
presented schematically in Fig. \ref{fig:Fig1} (b) and mathematically below are not
unrealistic for a practical on-chip device.

The heat balance between the resistor and the NIS structure can be
described on the level of a single electron tunnelling event between
N and S, accompanied by an exchange of energy with the junction's
resistive environment. Formally this can be done using the so-called
$P(E)$ theory developed for a tunnel junction embedded in an
electromagnetic environment \cite{devoret90,petheory}. Assuming that
for both electrodes $i=1,2$ the (normalized) density of states
$n_i(E)$ is symmetric around the Fermi level $E=0$ and that the
corresponding energy distributions satisfy $f_i(E)=1-f_i(-E)$, we
find that the heat flux out of $1$ upon tunnelling from electrode
$1$ to $2$ is given by
\begin{eqnarray} \label{enrate1}
\dot{Q}_{1} = && \frac{2}{e^2R_{\rm T}}\int \int dE' dE
n_1(E')n_2(E)\times \nonumber \\
&& E'f_1(E')[1-f_2(E)]P(E'-E).
\end{eqnarray}
Here, $P(E'-E)$ is the probability density of emitting energy $E'-E$ to the environment
in a tunnelling process from $1$ to $2$. It can be calculated for a particular
environment at a temperature $T_{\rm R}$, once the dissipative part $\Re\mbox{e}
Z(\omega)$ of its impedance at frequency $\omega /2\pi$ is known \cite{petheory}. The
theory is perturbative in tunnel conductance, therefore the tunnel resistance $R_{\rm T}$
should be of the order of the resistance quantum, $R_{\rm K} =h/e^2$, or higher. Here, we
have neglected the Joule dissipation in the normal metal itself, which is usually
justified because N can have a very low resistance as compared to that of the tunnel
junction.

As a warm-up exercise let us look at the simplest system, where the tunnel junction is of
type NIN, i.e., both sides are normal metals. Then $n_i(E) \equiv 1$ for both electrodes;
we furthermore assume them to be kept at equal temperature $T_N$ and characterized by
equilibrium Fermi distributions $f_i(E)=(1+e^{E/k_{\rm B}T_{\rm N}})^{-1}$. This example
closely resembles the original set-up of Johnson and Nyquist \cite{kogan}, where thermal
noise acts between two resistors. At high temperatures, $k_{\rm B}T_{\rm R,N} \gg \hbar
(RC)^{-1}$, we then obtain, using Eq. \eqref{enrate1}, the total heat flux out of the
junction as $\dot{Q} \equiv \dot{Q}_1 + \dot{Q}_2 \simeq k_{\rm B}(T_{\rm N}-T_{\rm
R})(R_{\rm T}C)^{-1}$. This power is shared equally by both electrodes due to symmetry.
The result for $\dot{Q}$ is consistent with the expectation that heat flows from hot to
cold.

The symmetry of the previous example is vitally broken in the NIS junction that we focus
on here. In particular, we have $n_1(E)=1$ for N as before, but the BCS density of states
in S, $n_2(E)=0$ for $|E| < \Delta$ and $n_2(E)=|E|/\sqrt{E^2-\Delta^2}$ for $|E| >
\Delta$, makes this system behave as an energy selective entity in the spirit of
Maxwell's demon. Here, $\Delta$ is the energy gap of the superconductor. We again assume
standard Fermi distributions, but possibly at different temperatures:
$f_1(E)=(1+e^{E/k_{\rm B}T_{\rm N}})^{-1}$ for N and $f_2(E)=(1+e^{E/k_{\rm B}T_{\rm
S}})^{-1}$ for S.

Let us first discuss a particular limit where illustrative
analytical results can be obtained. We consider a very resistive
environment, such that $\pi \frac{R}{R_{\rm K}}\frac {k_{\rm
B}T_{\rm R}}{E_{\rm C}} \gg 1$, where $E_{\rm C}\equiv
\frac{e^2}{2C}$ is the charging energy. Then $P(E)$ assumes a simple
Gaussian form: $P(E) =(2\pi \sigma)^{-1/2}\exp(-\frac{(E-E_{\rm
C})^2}{2\sigma})$, where the width is given by $\sigma=2k_{\rm
B}T_{\rm R}E_{\rm C}$ \cite{petheory}. We now assume that this width
is small compared to the superconducting gap $\Delta$. The
integration over energy in (\ref{enrate1}) then involves $E,E' \agt
\Delta \gg k_{\rm B}T_{\rm N,S}$ and we may approximate the Fermi
distributions by their Boltzmann-like exponential tails. In this
limit, setting further $T_{\rm S}=T_{\rm N}$ for simplicity, we have
\begin{eqnarray} \label{QNISnet}
&& \dot{Q}_{\rm N} \simeq \frac{\sqrt{2\pi \Delta k_{\rm B}T_{\rm
N}}}{e^2 R_{\rm T}}\Delta e^{-\Delta/k_{\rm B}T_{\rm N}}\times
\nonumber \\&&\{ [1- (2 \frac{T_{\rm R}}{T_{\rm N}}-1)\frac{E_{\rm
C}}{\Delta} ]e^{(\frac{T_{\rm R}}{T_{\rm N}}-1)E_{\rm C}/k_{\rm
B}T_{\rm N}}+ \frac{E_{\rm C}}{\Delta} -1\}.
\end{eqnarray}
One can similarly estimate the heat flux from the superconductor:
\begin{equation} \label{PNISnet}
\dot{Q}_{\rm S} \simeq \frac{\sqrt{2\pi \Delta k_{\rm B}T_{\rm
N}}}{e^2 R_{\rm T}}\Delta e^{-\Delta/k_{\rm B}T_{\rm
N}}[1-e^{(\frac{T_{\rm R}}{T_{\rm N}}-1)E_{\rm C}/k_{\rm B}T_{\rm
N}}].
\end{equation}
The expressions \eqref{QNISnet} and \eqref{PNISnet} vanish when
$T_{\rm R} = T_{\rm N}$, as expected. The non-trivial result is that
on heating the environment to temperatures $T_{\rm R}
> T_{\rm N}$, N tends to cool down, i.e., $\dot{Q}_{\rm N} > 0$,
and S is heated, $\dot{Q}_{\rm S} < 0$.
On the other hand, for $T_{\rm R} < T_{\rm N}$ the heat flux is
reversed: S tends to cool down and N to warm up. Also in this regime
the heat flux between the junction and the resistor is unevenly
distributed among N and S. Such cooling of S never occurs in a
conventional voltage biased NIS refrigerator \cite{giazotto06}.

If one employs \eqref{QNISnet} to find the optimum $T_{\rm R}/T_{\rm N}$ where the
cooling power of N is maximal, one finds $T_{\rm R}/T_{\rm N}\simeq \Delta/2E_{\rm C}$
for $E_{\rm C}/\Delta \ll 1$. Although this is the right order of magnitude, the
numerical results presented below show that the actual value of the ratio $T_{\rm
R}/T_{\rm N}$ is approximately twice higher. The approximation used above for the Fermi
functions leads to a counterbalance of the cooling effect as the Boltzmann tails grow
exponentially at negative energies. A better estimate can be obtained by retaining the
full Fermi function $f_1(E')$ and linearizing the exponent of the Gaussian $P(E'-E)$
around $E' =0$ where $f_1$ steeply drops. We then arrive at
\begin{eqnarray} \label{cp}
&& \dot{Q}_{\rm N} \simeq \frac{\pi^3 (k_{\rm B}T_{\rm
N})^2}{2e^2R_{\rm T}\sqrt{1+E_{\rm C}/\Delta}}\times \nonumber \\
&& [(\Delta/E_{\rm C}+1)T_{\rm N}/T_{\rm
R}-1]e^{-\frac{\Delta^2}{4k_{\rm B}T_{\rm R}E_{\rm C}}(1+E_{\rm
C}/\Delta)^2}.
\end{eqnarray}
This expression predicts that for $k_{\rm B}T_{\rm N}, E_{\rm C}\ll\Delta$ the optimal
point of cooling indeed lies at $T_{\rm R}/T_{\rm N} \simeq \Delta/E_{\rm C}$. In its
range of validity this optimum as well as the overall behavior described by Eq.
\eqref{cp} are consistent with the numerical results, as will be seen below.
\begin{figure}
\begin{center}
\includegraphics[width=0.5
\textwidth]{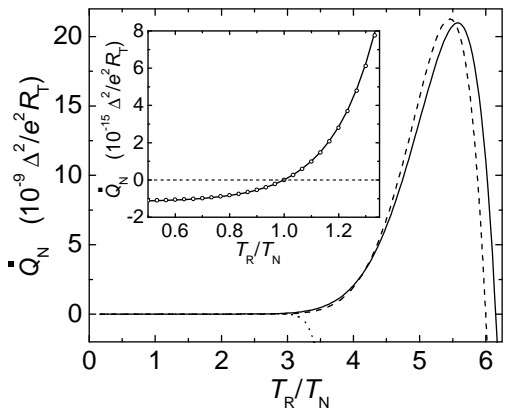} \caption{Calculated cooling power
$\dot{Q}_{\rm N}$ at $k_{\rm B}T_{\rm N}/\Delta =0.03$ as a function
of the temperature bias $T_{\rm R}/T_{\rm N}$. Here we have assumed
that $R\gg R_{\rm K}$, $T_{\rm S} =T_{\rm N}$, and $\Delta/E_{\rm C}
= 5$. In the main frame the solid line is the result of the exact
numerical calculation, and the dashed one is the analytic expression
of Eq. \eqref{cp} based on linearizing the exponent of the Gaussian
distribution over the relevant energy interval. The dotted line
diving to negative values around $T_{\rm R}/T_{\rm N}\simeq 3$ is
the result of Eq. \eqref{QNISnet}. This approximation works well,
contrary to Eq. \eqref{cp}, at small temperature biases, as shown by
the inset: the solid line is the exact result and the approximation
\eqref{QNISnet} is shown by open dots.} \label{fig:Fig2}
\end{center}
\end{figure}

In order to picture the characteristics of the system in a
dissipative environment of arbitrary resistance $R$, we have
performed numerical calculations. The usual environment theory of
single-electron tunneling was employed here also. In all the
numerical calculations the ideal junction was coupled to a parallel
$RC$ circuit. In Fig. \ref{fig:Fig2} we show a comparison of the
exact numerical results for $R\gg R_{\rm K}$ and the analytical
approximations of Eqs. \eqref{QNISnet} and \eqref{cp}. We see in the
main frame that at low temperatures, in this case at $k_{\rm
B}T_{\rm N}/\Delta =0.03$, the approximation \eqref{cp} works quite
well over a broad range of temperature biases, whereas the Boltzmann
approximation \eqref{QNISnet} fails at high values of $T_{\rm
R}/T_{\rm N}$. Yet, around zero temperature bias the latter
approximation works perfectly as demonstrated by the inset of Fig.
\ref{fig:Fig2}.

\begin{figure}
\begin{center}
\includegraphics[width=0.5
\textwidth]{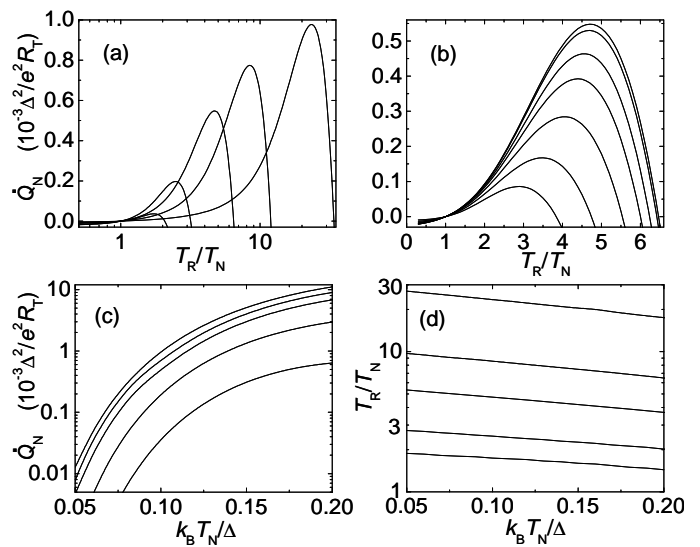} \caption{Heat flux out from the normal island.
In (a) and (b) we assume that $k_{\rm B}T_{\rm N}/\Delta =0.1$ and
that the superconductor temperature $T_{\rm S} =T_{\rm N}$. The
cooling power is plotted as a function of the ratio $T_{\rm
R}/T_{\rm N}$. In (a) the environment resistance is $R \gg R_{\rm
K}$. Different curves, with maxima from left to right, correspond to
$\Delta/E_{\rm C} = 1,2,5,10$ and $30$. In (b) $\Delta/E_{\rm C} =
5$, and the resistance varies from top to bottom as $R/R_{\rm
K}=\infty,10,2,1,0.5,0.25,$ and $0.125$. In (c) we plot the maximum
cooling power and in (d) the relative temperature of the resistor at
this optimum point as functions of the N island temperature. In (c)
and (d) the curves from top to bottom are for $\Delta/E_{\rm C} =
30, 10, 5, 2,$ and $1$.}\label{fig:Fig3}
\end{center}
\end{figure}

A set of numerical results under representative conditions is
collected in Fig. \ref{fig:Fig3}. In (a) we see the influence of
$E_{\rm C}$ on the performance of the system. Here the Gaussian
approximation of $P(E)$ was used. The maximum cooling occurs indeed
at $T_{\rm R}/T_{\rm N}\simeq \Delta/E_{\rm C}$, and the value at
the maximum grows a little with increasing $\Delta/E_{\rm C}$. In
(b) the Gaussian approximation was abandoned. Under reduced $R$ the
$P(E)$ function transforms from a broad Gaussian (width $2k_{\rm
B}T_{\rm R}E_{\rm C}$) centered around $E_{\rm C}$ towards a
delta-function around $E=0$. According to Eq. \eqref{enrate1} this
evolution weakens the refrigeration effect. In Fig. \ref{fig:Fig3}
(b) we see that it is indeed essential to have a relatively large
$R$ in order to sustain the effect, although some cooling can be
observed also down to $R/R_{\rm K} \sim 0.1$. We also see that with
high environment resistances the cooling characteristics approach
the result of a Gaussian environment as should be the case. In (c)
and (d) we show the quantitative results for the heat flux
$\dot{Q}_{\rm N}$ and the optimum temperature bias, respectively, as
a function of $T_N$ in a highly resistive environment.

Although the discussion above demonstrates counter-intuitive heat
fluxes in the system, it is straightforward to verify that they do
not contradict the second law of thermodynamics. For instance,
consider the resistor and the NIS junction to form an isolated
system such that, referring to Fig. \ref{fig:Fig1} (b), $P_{\rm
ext}=0$ and that couplings to the phonon bath vanish. Initially
brought to temperatures $T_{\rm R}$ and $T_{\rm N}$ ($=T_{\rm S}$
for simplicity), the system then starts to relax towards a common
temperature. For a small temperature difference $\Delta T \equiv
T_{\rm R} -T_{\rm N}$, we have $\dot{Q}_{\rm N}+\dot{Q}_{\rm
S}\simeq -\left[\frac{\sqrt{2\pi \Delta k_{\rm B}T_{\rm N}}}{e^2
R_{\rm T}} (2+\frac{E_{\rm C}}{k_{\rm B}T_{\rm N}})E_{\rm C}
e^{-\Delta/k_{\rm B}T_{\rm N}}\right]\frac{\Delta T}{T_{\rm N}}$
according to Eqs. \eqref{QNISnet} and \eqref{PNISnet}. This is the
heat flux between the junction and the resistor, and it is always
directed from hot to cold thereby ensuring positive entropy
production.

Some of the results above can be obtained approximately by straightforward classical
estimates. In particular, we may consider an ordinary NIS junction with large resistance
$R_{\rm T}$ and evaluate its cooling power when biased by a fluctuating voltage with
vanishing mean value and Gaussian variance $\langle (eV)^2 \rangle \equiv \sigma =
2k_{\rm B}T_{\rm R}E_{\rm C}$. In the limit of low frequencies (small $E_{\rm C}$), we
may make a quasi-stationary averaging over the fluctuations, such that the expected
cooling power of N is $\langle \dot{Q}_{\rm N}\rangle \simeq \int p(eV)\dot{Q}_{\rm N}
d(eV)$, where $p(eV)=(2\pi \sigma)^{-1/2}e^{-(eV)^2/(2\sigma)}$ is the distribution of
fluctuations and $\dot{Q}_{\rm N}(eV)$ is the cooling power of the NIS junction at a
static bias voltage $V$. For $|eV| \gg k_{\rm B}T_{\rm N}$, one obtains $\dot{Q}_{\rm
N}(eV)\simeq \frac{\sqrt{\pi\Delta k_{\rm B}T_{\rm N}/2}}{e^2R_{\rm T}}(\Delta -
|eV|)e^{-(\Delta-|eV|)/k_{\rm B}T_{\rm N}}$ \cite{anghel01}. Performing the averaging
yields then $\langle \dot{Q}_{\rm N}\rangle \simeq \frac{\sqrt{2\pi\Delta k_{\rm B}T_{\rm
N}}}{e^2R_{\rm T}}\Delta e^{-\Delta/k_{\rm B}T_{\rm N}}(1-2\frac{T_{\rm R}}{T_{\rm
N}}\frac{E_{\rm C}}{\Delta})e^{\frac{T_{\rm R}}{T_{\rm N}}E_{\rm C}/k_{\rm B}T_{\rm N}}$.
This result resembles very closely that of Eq. \eqref{QNISnet}, in particular for $T_{\rm
R} > T_{\rm N}$. Yet the classical result above neglects the backflow of heat from the
junction to the resistor, which is an important contribution especially when $T_{\rm R}
\lesssim T_{\rm N}$.
\begin{figure}
\begin{center}
\includegraphics[width=0.5
\textwidth]{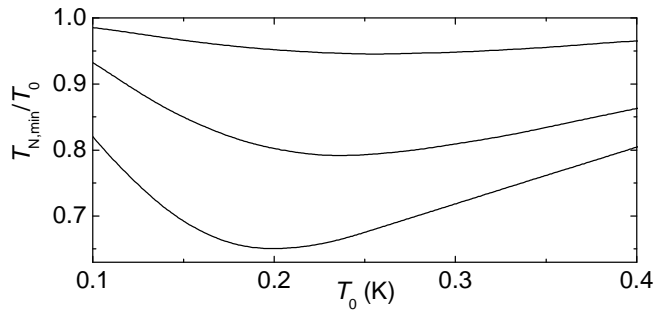} \caption{Temperature reduction in N in a
realistic device at the optimized resistor temperature. The
resistance is assumed to be high, $R \gg R_{\rm K}$. The calculation
has been performed for aluminium as a superconductor with $\Delta =
200$ $\mu$eV. The other parameters are: $\Delta/E_{\rm C}=5$,
$\Sigma = 1\cdot 10^9$ WK$^{-5}$m$^{-3}$, and
$\Omega=1\cdot10^{-21}$ m$^3$. From top to bottom the curves
correspond to $R_{\rm T} = 100, 10$ and $1$ k$\Omega$.
}\label{fig:Fig4}
\end{center}
\end{figure}

Next we discuss the possible experimental demonstration of the
presented effect. The basic concept can be implemented for instance
in a standard on-chip configuration by employing common N and S
metals like copper and aluminium, respectively. The resistance can
be formed using a strip of resistive metal such as chromium or a
metallic alloy. Figure \ref{fig:Fig4} demonstrates the calculated
temperature reduction, $T_{\rm N}/T_0$, of the N island assuming
that the superconductor is well thermalized at the phonon bath
temperature $T_0$. The results are shown specifically for the case
of aluminium as a superconductor ($\Delta = 200$ $\mu$eV, transition
temperature $T_{\rm C}\simeq 1$ K). We have chosen the ratio
$\Delta/E_{\rm C}=5$, which, on one hand corresponds to realistic
junction parameters, and on the other hand allows maximum cooling to
occur at not excessively high resistor temperatures: the largest
temperature reduction of almost 40\% occurs at $T_0 \simeq 0.2$ K
corresponding to $T_{\rm R}\simeq 0.7$ K. The specific material
parameter for the electron-phonon coupling in the normal metal was
chosen to be $\Sigma = 1\cdot 10^9$ WK$^{-5}$m$^{-3}$
\cite{giazotto06}. The N island volume of $\Omega=1\cdot10^{-21}$
m$^3$ can be achieved by a standard process.

In the analysis above the charging energy $E_{\rm C}$ is due to the
parallel connection of the junction capacitance and an optional
shunt capacitance. We have seen that the optimum cooling results do
not depend particularly strongly on $E_{\rm C}$, as long as $E_{\rm
C}< \Delta$ holds. Some improvement in performance can, however, be
obtained by decreasing $E_{\rm C}$, see Fig. \ref{fig:Fig3}, because
the parallel capacitance filters out the harmful high frequency tail
of the noise spectrum. Yet the ratio $T_{\rm R}/T_{\rm N}\sim
\Delta/E_{\rm C}$ reflects the fact that with large shunting
capacitance one needs to have a hotter noise source to induce
sufficiently strong fluctuations. From the practical point of view
this is not advantageous, at least in a conventional on-chip
solution, because the high temperature of the environment leads to
parasitic heating of N via the phonons of the substrate. Therefore,
in a practical realization the choice of $C$ is a trade-off.

We believe that the presented parallel $RC$ environment can be realized almost exactly as
long as $C$ is kept small, say on the level of few fF arising from the junction itself.
Then the series inductance of a circuit of sub-100 $\mu$m dimensions can be neglected,
because the corresponding $LC$-frequency remains high as compared to the frequency band
of thermal radiation at sub-K temperatures.

Finally, the effect can also be realized in a symmetric SINIS configuration, instead of
using a single NIS junction with an NS Andreev mirror. There the same theoretical
analysis can be carried over by replacing the environment resistance by an effective
resistance $R/4$, and the capacitance by $2C$, where $R$ is the true environment
resistance and $C$ is the capacitance of one junction. To take into account the possible
charging effects on the island, a more involved analysis needs to be performed, however.

In summary, we have discussed a Brownian refrigerator of electrons, which offers an
illustrative example of Maxwell's demon in the form of a tunnel junction with a
superconducting energy gap. Its operation is based on building blocks whose
characteristics and implementation are well known. We expect that it yields a substantial
temperature reduction in a straightforward on-chip realization.

We thank M. B\"uttiker, M. Meschke, T. Heikkil\"a and O.-P. Saira
for discussions and M. Helle for checking some of the numerical
results. NanoSciERA project "NanoFridge" of EU and Institut
universitaire de France are acknowledged for financial support.

\end{document}